\RequirePackage{fix-cm}
\documentclass[smallextended]{svjour3}       
\smartqed  
\usepackage{graphicx}
\usepackage{amstext}
\usepackage{amsmath}
\begin{document}

\title{Effect of nuclear deformation on Gamow-Teller strength distributions of Hg isotopes}


\author{Jameel-Un Nabi \and
       Muhammad Riaz \and
       Tuncay Bayram \and
       Muhammad Azaz
}

\institute{Jameel-Un Nabi \at
	University of Wah, Wah Cantt 47040,  Punjab, Pakistan.\\
	GIK Institute of Engineering Sciences and Technology, Topi 23640, KP, Pakistan.\\
	\email{jameel@uow.edu.pk}           %
	\and
	Muhammad Riaz\at
	Department of Physics, University of Jhang, Jhang, Punjab Pakistan\\ 
	University of Wah, Wah Cantt 47040, Punjab, Pakistan\\ 
	\and
	Tuncay Bayram \at
	Karadeniz Technical University, Turkey\\  
	\and
	Muhammad Azaz \at
	University of Wah, Wah Cantt 47040, Punjab, Pakistan\\           
}

\date{Received: date / Accepted: date}

\maketitle

\begin{abstract}
Recent studies \cite{Boi15,Mor06} predicted the sensitivity of the Gamow-Teller (GT) strength distributions on nuclear deformation in neutron-deficient Hg isotopes. Motivated by this work we investigate nuclear ground state properties and GT strength distributions for neutron-deficient Hg isotopes ($^{177-193}$Hg).   The nuclear deformation ($\beta(E_2)$) values were calculated using the Relativistic Mean Field (RMF) model. The RMF approach with different density-dependent interactions, DD-ME2 and DD-PC1, was used to compute nuclear shape parameters. The computed deformation values were later used within the framework of deformed proton-neutron quasi-random phase approximations (pn-QRPA) model, with a separable interaction, to calculate the allowed GT strength distributions for these Hg isotopes. Our calculations validate the findings of \cite{Boi15} and confirmed the  effect of deformation on GT strength distributions. Our study may further provide a complementary signature for nuclear shape isomers. Noticeable differences are highlighted between the current and previous calculations. 
The calculation of \cite{Boi15}  shows that $^{177-182}$Hg posses prolate and $^{184-196}$Hg have oblate shapes. Our calculation, on the other hand, 
predicts prolate shape for $^{177-188}$Hg and oblate shape for $^{189-193}$Hg isotopes. We finally analyze the effect of deformation on the calculated $T_{1/2}$ for  the selected Hg isotopes. Our half-life calculations do not validate the findings of \cite{Mor06} that $T_{1/2}$ values are not good enough observables to study deformation effects.  
\end{abstract}

\noindent{\rm Keywords}: Gamow-Teller strength; nuclear deformation; half-life; deformed pn-QRPA; nuclear structure.

\section{Introduction}
\label{intro}
Shape coexistence and shape transition play a crucial role for understanding the low-lying nuclear structure properties and characterizing the isotopes in neutron deficient region~\cite{Mor06,Bre14}. Shape coexistence may be regarded as the outcome of two opposing forces inside the atomic nucleus. The first interaction has a stabilizing effect (due to closed shells and subshells) and tends to preserve spherical shape of the nucleus. The second force is the residual interaction between the nucleons and supports deformation of the nucleus.  The concept of shape coexistence \cite{Mor56}  exists in both light and heavy nuclei. Theoretical models still face challenges in providing a sound and microscopic explanation of the occurrence of shape coexistence. The shape coexistence of neutron-deficient nuclei was experimentally investigated using the $\gamma$-rays in coincident of $\alpha$ emission during the specified nuclear reactions. Nuclear reaction products of the fusion evaporation reactions led to $\alpha-$decay, containing very important information at this stage for shape coexistence.    
$^{180-190}$Hg isotopes were found to be oblate in the stationary state and prolate within excitation energy of 1 $MeV$ \cite{Mor06}. According to Bonn \cite{Bon72}, quadrupole deformation exist in the region where nucleon shells are not completely filled, and this effect of shape coexistence was observed by isotopic shift experiments in $^{183,185}$Hg.

It was found by mean field theory that, relative energy of the different nuclear shape is highly sensitive to calculation of pairing effect for neutron-deficient nuclei including Hg and Pb. This effect is valid both in  relativistic \cite{Nik02,Yos94,Yos97} and non-relativistic \cite{Taj93} calculations.   
In order to study the electromagnetic properties of Hg isotopes, Coulomb-excitation experiments were performed which further confirmed the shape coexistence in the light even-even Hg isotopes \cite{Bre14}. 
 
To explain the shape coexistence at lower energies, two different theoretical approaches (spherical shell model and mean field) were used \cite{Hey11}. The first recipe is to start from a spherical mean field, incorporate high energy states as multiple particle-hole excitations. In this approach description of shape coexistence phenomena may be met by incorporating at least two major shells. In the second approach, one uses the two-body nucleon-nucleon force to obtain single-particle states in a self-consistent manner. Pairing correlations are formulated in Hartree-Fock-Bogoliubov
(HFB) theory to simultaneously optimize both the mean
single-particle field and the pairing  properties
in nuclei.

Moreno et al.~\cite{Mor06} studied the effect of nuclear deformation on the energy distributions of the GT strength in
neutron-deficient Hg, Pb, and Po even isotopes. The calculation was performed using a self-consistent deformed quasiparticle Skyrme Hartree–Fock basis. The SLy4 \cite{Cha98} interaction was used for the calculation being one of the most successful Skyrme force. It was concluded that whereas the GT strength distributions got altered as deformation values changed, the $\beta$-decay half-lives and the summed strengths did not change significantly. More recently a pn-QRPA calculation of GT strength distributions of Hg isotopes was performed \cite{Boi15} . The calculations displayed a shape transition in agreement with the measured data and described the decay properties in terms of deformation. Motivated by this calculation we present here a similar investigation of GT strength distributions and $\beta$-decay half-lives of Hg isotopes using the deformed pn-QRPA model with a separable schematic interaction. We selected seventeen (17) isotopes of Hg (A=177-193). These isotopes are few of the the best candidates for experimental investigation of $\beta-$decay, due to their large $T_{1/2}$ and associated Q-values.  The nuclear deformations for selected Hg isotopes were calculated using the Relativistic Mean-Field (RMF) model using two different interactions and then later used for calculation of GT strength distributions employing the deformed  proton-neutron quasi-random phase approximations (pn-QRPA) model with a separable schematic interaction. Our findings support the conclusion of earlier calculation \cite{Boi15} that  $\beta-$decay properties of neutron-deficient Hg isotopes get altered by changing the values of  nuclear deformation. 

In next section we present the formalism of RMF and deformed pn-QRPA models, followed by the results discussed in Section~3. We finally conclude our investigations in Section~4.
\section{Formalism}
The RMF model with two different interactions was employed to compute the deformation parameter values for selected $^{177-193}$Hg isotopes. The calculated nuclear deformation parameters were later used to compute GT strength distributions and $\beta$-decay $T_{1/2}$ for  $^{177-193}$Hg using the deformed pn-QRPA model. Below we briefly describe the formalism adopted in the two nuclear models to perform our necessary calculations.

\subsection{Relativistic Mean Field (RMF) Model}
In the RMF model, the nucleons are considered to  oscillate independently in the mean field produced by the exchange of mesons and photons~\cite{walecka1974,boguta1977,ring1996,lalazissis1999,vretenar2005,bayram13a,Bay13,bayram19}. By means of self-interaction of mesons and density-dependent meson-nucleon couplings, a few types of non-linear RMF model versions could be found (for more details and references see Ref. ~\cite{meng2006}). In the present work, only $\sigma$, $\omega$ and $\rho$-mesons are considered. A brief description of the RMF model-based density-dependent meson-nucleon couplings is given in this subsection. The RMF model starts with the phenomenological relativistic Lagrangian density, which is given by
\begin{equation}
	\begin{split}
		L &= \bar{\Psi}(i\gamma\partial-m)\Psi+\frac{1}{2}(\partial\sigma)^2-\frac{1}{2}m_\sigma\sigma^2-\frac{1}{4}{\bf \Omega}_{\mu\nu}{\bf \Omega}^{\mu\nu}+\frac{1}{2}m^2_\omega\omega^2  \\
		&-\frac{1}{4}\overrightarrow{\bf R}_{\mu\nu}\overrightarrow{\bf R}^{\mu\nu}+\frac{1}{2}m^2_\rho\overrightarrow{\rho}^2-\frac{1}{4}{\bf F}_{\mu\nu}{\bf F}^{\mu\nu}-g_\sigma\bar{\Psi}\sigma\Psi \\
		&-g_\omega\bar{\Psi}\gamma.\omega\Psi-g_\rho\bar{\Psi}\gamma.\vec{\rho}\vec{\tau}\Psi-e\bar{\Psi}\gamma.A\frac{1-\tau_3}{2}\Psi
	\end{split}
	\label{lagrangian}
\end{equation}
where $m$, $m_\sigma$, $m_\omega$, $m_\rho$ are the mass of nucleon, $\sigma$, $\omega$ and $\rho$ mesons, respectively. The  $g_\sigma$, $g_\omega$ and $g_\rho$ are the coupling constants of the mesons. The interaction strengths of nucleons with meson fields are represented by these coupling constants. Vector field tensors in this equation are given by 
\begin{equation}
	\begin{split}
		{\bf \Omega}^{\mu\nu}=\partial^\mu\omega^\nu-\partial^\nu\omega^\mu, \\
		\overrightarrow{\bf R}^{\mu\nu}=\partial^\mu\overrightarrow{\rho}^\nu-\partial^\nu\overrightarrow{\rho}^\mu, \\
		{\bf F}^{\mu\nu}=\partial^\mu A^\nu-\partial^\nu A^\mu \\
	\end{split}
\end{equation}
where the photon field, $A^\mu$, is related with electromagnetic interaction due to protons. 
In Eq.~(\ref{lagrangian}),  vectors in three-dimensional space are shown in bold symbols while vectors in isospin space are  indicated with arrows. The unknown
meson masses and coupling constants are fitted from a few experimental data. 
The meson-nucleon vertex functions can be fixed-on
by tuning the parameters of the assumed phenomenological density
dependence of the meson-nucleon couplings to generate the
properties of symmetric and asymmetric nuclear matter and finite
nuclei. For details of RMF formalism we refer to ~\cite{lalazissis2005}.

The Lagrangian is invariant under parity transformation and, since only the solutions with defined parity are considered, the expectation value of the pseudoscalar pion field vanishes in the Hartree Fock approximation. By applying the classical variational principle to the relativistic Lagrangian density given in Eq.~(\ref{lagrangian}), the RMF equations can be obtained. The coupled equations that are obtained are Dirac equation with the potential terms for the nucleons and the Klein$-$Gordon-like equations with sources for mesons and photon. The wave functions can be extended in terms of a spherical, axially, and triaxially symmetric harmonic oscillator potential to solve equations individually ~\cite{niksic2014}.  In this work, we have used axially symmetric RMF calculations to produce ground state quadrupole deformation parameter, $\beta_2$, for prolate and oblate shape configurations of $^{176-193}$Hg isotopes  by following the recipe of Ref .~\cite{niksic2014}. Two different versions of density-dependent RMF interactions, DD-ME2~\cite{lalazissis2005} and DD-PC1~\cite{niksic2008},
were used in our calculations. 12 shells of neutrons and protons were taken into account and no problem of convergence was reported in our iterative calculations. 

\subsection{The Deformed pn-QRPA Model}

For calculation of GT strength distributions, the Hamiltonian of the deformed pn-QRPA was taekn as 
\begin{equation} \label{H}
	H^{QRPA} = H^{sp} + V^{pair} + V^{pp}_{GT} + V^{ph}_{GT},
\end{equation}
where $H^{sp}$, $V^{pair}$, $V_{GT}^{pp}$ and $V_{GT}^{ph}$ are single particle Hamiltonian, nucleon-nucleon pairing interaction (for which the BCS approximation was considered), \textit{pp} and \textit{ph} GT interaction strengths, respectively. The Nilsson model \cite{nil55} was employed to calculate the wave functions and energies in a deformed basis. The oscillator constant was computed using $\hbar\omega=41A^{1/3}$. The Nilsson-potential parameters were chosen from Ref.~\cite{Rag84}. The pairing gaps were calculated employing $\triangle_n=\triangle_p={12/\sqrt A}$  MeV~\cite{Har09}.
$Q$-values were calculated from the mass excess values taken from the mass compilation of Audi et al. \cite{Aud17}. 
The spherical basis was transformed to the (axial-symmetric) deformed basis using
\begin{equation}\label{df}
	d^{\dagger}_{m\alpha}=\Sigma_{j}D^{m\alpha}_{j}c^{\dagger}_{jm},
\end{equation}
where $d^{\dagger}$ and $c^{\dagger}$ are particle creation operators in the deformed and spherical basis, respectively. The matrices $D^{m\alpha}_{j}$ were obtained by diagonalizing the Nilsson Hamiltonian.\\ For pairing within BCS approximation, a constant pairing force was applied and a quasi-particle (q.p) basis was introduced:
\begin{equation}\label{qbas}
	a^{\dagger}_{m\alpha}=u_{m\alpha}d^{\dagger}_{m\alpha}-v_{m\alpha}d_{\bar{m}\alpha}
\end{equation}

\begin{equation}
	a^{\dagger}_{\bar{m}\alpha}=u_{m\alpha}d^{\dagger}_{\bar{m}\alpha}+v_{m\alpha}d_{m\alpha},
\end{equation}
where $\bar{m}$, $a^{\dagger}$ and $a$ represents the time reversed state of $m$, the q.p. creation and annihilation operator, respectively which comes in the RPA equation. The occupation amplitudes ($u_{m\alpha}$ and $v_{m\alpha}$) were computed using BCS approximation (satisfying $u^{2}_{m\alpha}$+$v^{2}_{m\alpha}$ = 1).\\
Within the deformed pn-QRPA framework, the GT transitions are described in terms of phonon creation and one describes the QRPA phonons as
\begin{equation}\label{co}
	A^{\dagger}_{\omega}(\mu)=\sum_{pn}[X^{pn}_{\omega}(\mu)a^{\dagger}_{p}a^{\dagger}_{\overline{n}}-Y^{pn}_{\omega}(\mu)a_{n}a_{\overline{p}}],
\end{equation}
where the indices $n$ and $p$ stand for $m_{n}\alpha_{n}$ and $m_{p}\alpha_{p}$, respectively, and differentiating neutron and proton single-particle states. The summation was taken over all proton-neutron pairs satisfying $\mu=m_{p}-m_{n}$ and $\pi_{p}.\pi_{n}$=1, with $\pi$ representing parity. In Eq.~(\ref{co}), $X$ and $Y$ represent the forward- and backward-going amplitudes, respectively, and are the eigenfunctions of the  RPA matrix  equation.  In the deformed pn-QRPA theory, the proton-neutron residual interactions work through \textit{ph} and \textit{pp} channels, characterized by interaction constants $\chi$ and $\kappa$, respectively. The $ph$ GT force can be expressed as
\begin{equation}\label{ph}
	V^{ph}= +2\chi\sum^{1}_{\mu= -1}(-1)^{\mu}Y_{\mu}Y^{\dagger}_{-\mu}\\
\end{equation}
with
\begin{equation}\label{y}
	Y_{\mu}= \sum_{j_{p}m_{p}j_{n}m_{n}}<j_{p}m_{p}\mid
	t_- ~\sigma_{\mu}\mid
	j_{n}m_{n}>c^{\dagger}_{j_{p}m_{p}}c_{j_{n}m_{n}},
\end{equation}
and the $pp$ GT force as
\begin{equation}\label{pp}
	V^{pp}= -2\kappa\sum^{1}_{\mu=
		-1}(-1)^{\mu}P^{\dagger}_{\mu}P_{-\mu}.
\end{equation}
with
\begin{eqnarray}\label{p}
	P^{\dagger}_{\mu}= \sum_{j_{p}m_{p}j_{n}m_{n}}<j_{n}m_{n}\mid
	(t_- \sigma_{\mu})^{\dagger}\mid
	j_{p}m_{p}>\times \nonumber\\
	(-1)^{l_{n}+j_{n}-m_{n}}c^{\dagger}_{j_{p}m_{p}}c^{\dagger}_{j_{n}-m_{n}},
\end{eqnarray}
where the remaining symbols have their usual meanings. 
Here, the different signs in \textit{ph} and \textit{pp} force reveal the opposite nature of these interactions i.e. \textit{pp} force is attractive while the \textit{ph} force is repulsive. The interaction constants $\chi$ and $\kappa$ were chosen in concordance with the suggestion given in Ref.~\cite{hom96}, following a $1/A^{0.7}$ relation. Authors in Ref.~\cite{hom96} performed a systematic study of the $\beta$-decay within the  framework of pn-QRPA and employed a schematic GT residual interaction. The \textit{ph} and \textit{pp} force were consistently included for both $\beta^+$ and $\beta^-$ directions, and their strengths were fixed as smooth functions of mass number A of nuclei in such a way that the calculation best reproduced the
observed $\beta$-decay properties of nuclei. Our computation further fulfilled the model independent Ikeda sum rule~\cite{Ike63}. The reduced transition probabilities for GT transitions from the QRPA ground state
to one-phonon states in the daughter nucleus were obtained as
\begin{equation}
	B_{GT} (\omega) = |\langle \omega, \mu ||t_{+} \sigma_{\mu}||QRPA \rangle|^2
\end{equation}
For further details and complete solution of Eq.~(\ref{H}), we refer to Hirsch et al. \cite{Hir91,Hir93}. 

The partial $\beta$-decay $T_{1/2}$ were calculated using the following relation
\begin{eqnarray}
	t_{p(1/2)} = \\ \nonumber
	\frac{D}{(g_A/g_V)^2f_A(Z, A, E)B_{GT}(E_j)+f_V(Z, A, E)B_F(E_j)},
\end{eqnarray}
where $E_j$ is the final state energy, $E$ = $Q_{EC}$ - $E_j$, $g_A/g_V$ (= -1.254)\cite{war94} represents ratio of axial vector to the vector coupling constant and D = $\frac{2\pi^3 \hbar^7 ln2}{g^2_V m^5_ec^4} = 6295 s$. Phase space factors $f_A(Z, A, E)$ ($f_V(Z, A, E)$) are the Fermi integrals for axial vector (vector) transitions. $B_{GT}$ and $B_F$ are the reduced transition probabilities for the GT and Fermi transitions, respectively. The $\beta$-decay half-life  was finally determined by summing up all transition probabilities to states in the daughter nucleus with excitation energies lying within the $Q_{EC}$ window:
\begin{equation}
	T_{1/2} = \left(\sum_{0 \le E_j \le Q_{EC}} \frac{1}{t_{p(1/2)}}\right)^{-1}
\end{equation} 
\section{Results and Discussion}
\begin{figure} 
	\hspace{-8.0em}	\includegraphics[width=9in,height=9in]{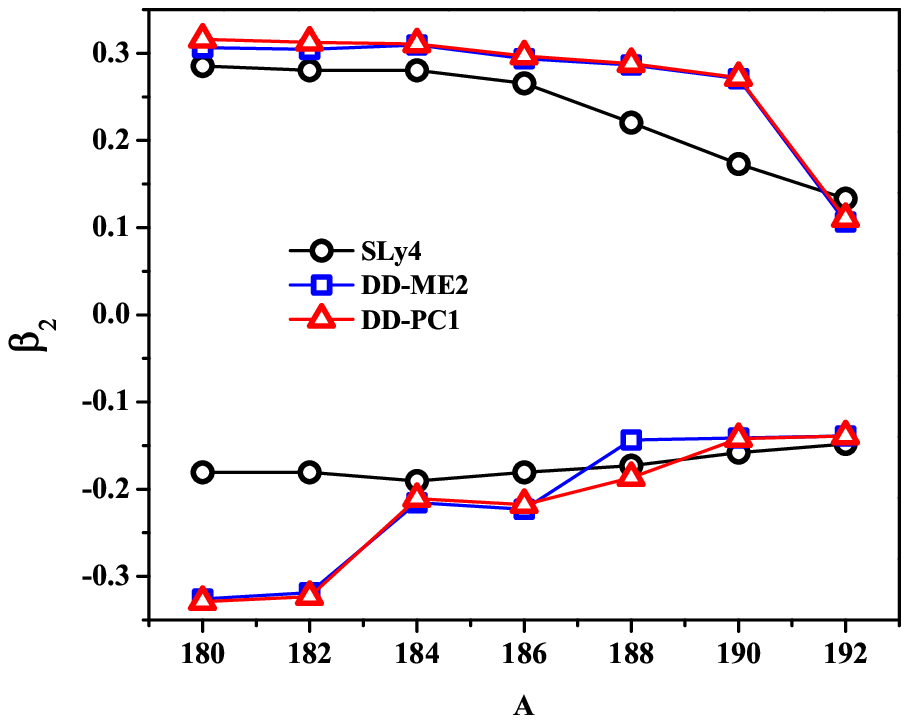}
	\caption{Comparison of the RMF calculated  nuclear deformation parameters using DD-ME2 and DD-PC1 interactions with those using the SLy4 interaction of Ref. \cite{Boi15}.}\label{fig1}
\end{figure}

In this section we discuss the computation of quadrupole deformation parameters ($\beta_2$) for $^{177-193}$Hg isotopes using the RMF model with DD-ME2 and DDPC1 interactions. Later we present the sensitivity of the calculated GT strength distributions and half-lives to the computed nuclear deformations. We further compare our calculated results with the experimental and previous calculation. \\
Fig.~\ref{fig1} shows the RMF model calculated nuclear deformation parameters using the DD-ME2 and DD-PC1 interactions for even-A isotopes of Hg. Shown also are the deformation values calculated using the SLy4 interaction of Ref. \cite{Boi15}.  The calculations show that as the neutron number increases the magnitude of the deformation parameter decreases and the nuclei tend to assume spherical shape. It is further noted that SLy4 interaction, in general, results in smaller magnitude of computed deformation parameters when compared with the RMF results.

Table.~\ref{Table 1} displays the computed quadrupole deformation parameter ($\beta_2$) values by using RMF model with the two different interactions (namely DD-ME2 and DD-PC1). Here we show the two minima predicted by the RMF model for the prolate and oblate shapes of Hg isotopes. The ground state deformation is the overall minimum and also shown in the table. For determining the ground state deformation parameter, the oblate or prolate configurations with the lowest total binding energy were selected. The oblate DD-ME2 deformation values are larger  by an average factor of 0.63 for the selected region of $^{177-193}$Hg isotopes than those predicted by FRDM calculation \cite{Mol12}.  The calculation of \cite{Boi15} concluded that the nuclei $^{174-182}$Hg posses  prolate  and $^{184-196}$Hg  oblate shapes. In contrast our calculation predict prolate shapes for  $^{177-188}$Hg  and oblate for $^{189-193}$Hg isotopes. As we move from  $^{177}$Hg towards $^{193}$Hg isotopes the computed deformation decreases by a factor of 1.07 which is consistent with FRDM deformations \cite{Mol12} having decrement with a factor of 0.99.

\begin{figure} 
	\hspace{-8.0em}	\includegraphics[width=9in,height=9in]{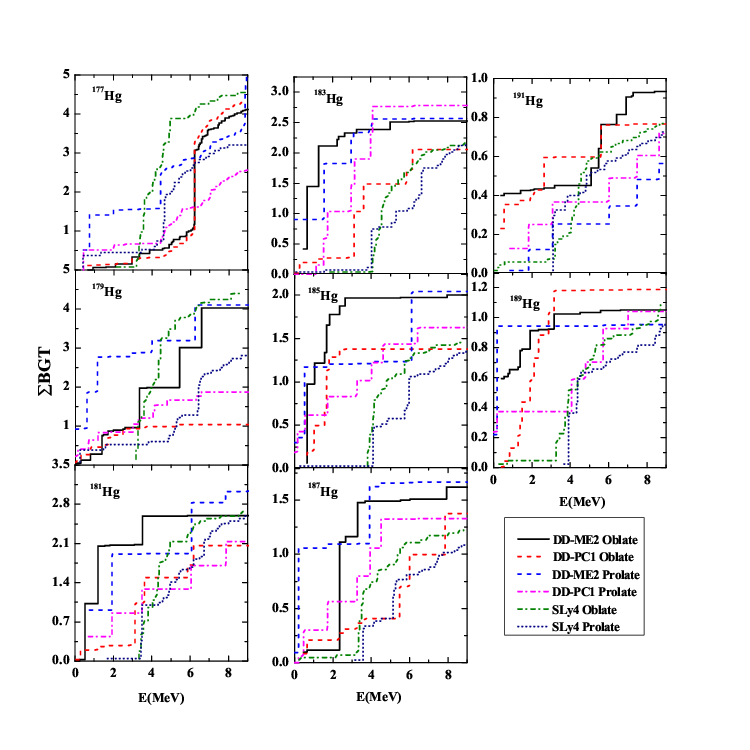}
	\caption{Calculated B(GT$_{+}$) strength
		distributions for odd-A Hg isotopes and comparison with previous calculation \cite{Boi15}.}\label{fig2}
\end{figure}

Next we investigate how the calculated GT strength distributions change by incorporating different values of nuclear deformation parameter in our pn-QRPA calculation. For doing the needful, the particle-hole ($\chi_{GT}$) and particle particle ($\kappa_{GT}$) force parameters were selected using the recipe in Ref.\cite{hom96}. The  $\chi_{GT}$ and $\kappa_{GT}$ forces were determined using the relation $\chi_{GT}=5.2/A^{0.7}$ and $\kappa_{GT}=0.58/A^{0.7}$. 
Fig.~\ref{fig2} shows the calculated GT strength distributions of odd-A isotopes $^{177-193}$Hg using the DD-ME2 and DD-PC1 interactions for oblate and prolate shapes. Our results are in  decent comparison with the SLy4 oblate and prolate calculations of Ref. \cite{Boi15}. The smaller GT values calculated using the Sly4 are attributed to the quenching of GT strength adopted in their calculation. In general, the DD-ME2 values of deformation resulted in bigger values of total GT strength up to 8 MeV in daughter nucleus.

\begin{figure} 
	\hspace{-5.0em}\includegraphics[width=7in,height=8in]{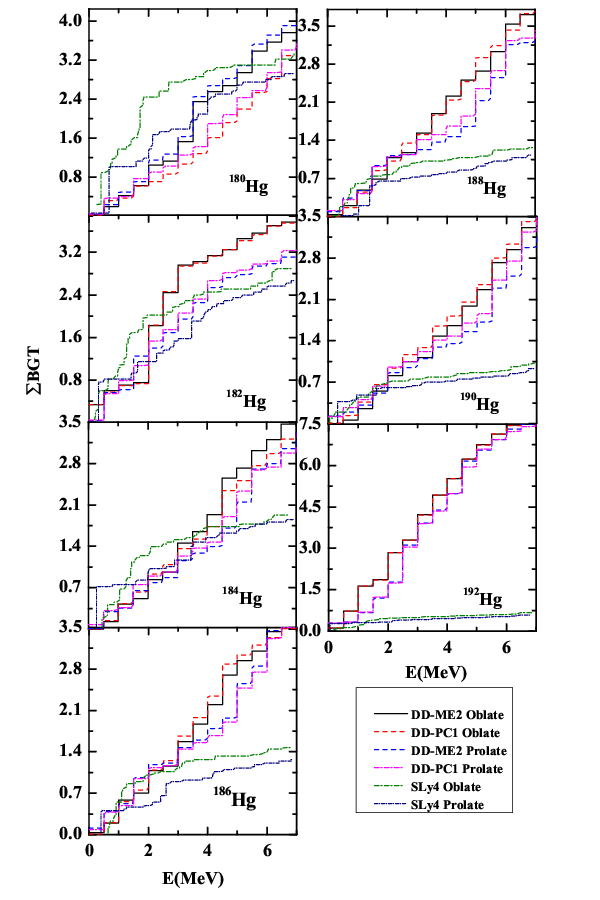}
	\centering \caption{Calculated B(GT$_{+}$) strength
		distributions for even-A Hg isotopes and comparison with previous calculation \cite{Boi15}.}\label{fig3}
\end{figure}

Fig.~\ref{fig3} shows a  similar comparison for even-A Hg isotopes.  Observation of GT strength distributions in case of even-A Hg isotopes shows that as the mass number increases the GT strength is shifted toward low-lying energy states in daughter. This effect may be attributed to the increasing number of neutrons as they approach the magic number 126. The profiles support the argument that GT distributions are characteristic of the nuclear shape and depend little on the details of the two-body force \cite{Boi15}. We further note that the  centroid values of the calculated strength distributions change significantly with the value of the deformation parameter as discussed below. 

Table.~\ref{Table 2} presents the centroids of calculated GT strength distributions for Hg isotopes using the different interactions employed in this work. The upper (lower) panel displays the computed centroid values for odd-A (even-A) Hg isotopes.  Similarly Table.~\ref{Table 3} shows the result for the computed total GT strengths. It is noted from Table.~\ref{Table 2} and Table.~\ref{Table 3} that the total GT strength changes marginally whereas the centroid values change substantially as we alter the nuclear shapes. Our results validate the findings of \cite{Mor06} that the summed strengths are not good observables to study deformation effects whereas GT strength distributions are.

\begin{figure} 
	\includegraphics[width=6in,height=5in]{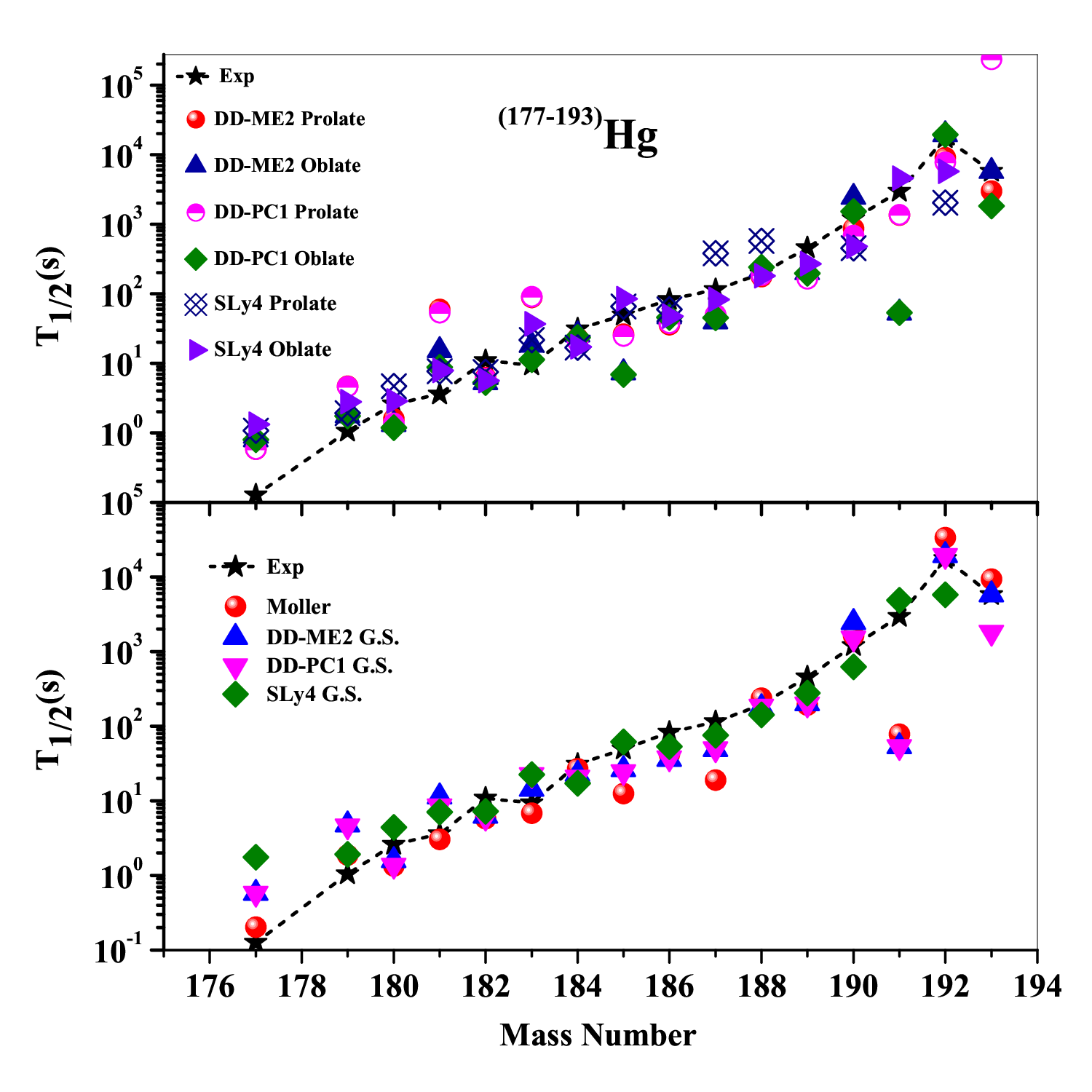}
	\centering \caption{Comparison of calculated and measured half-lives for $^{177-193}$Hg. For references see
		text.}\label{fig4}
\end{figure}

The measured and computed terrestrial half lives, using different values of nuclear deformation, for neutron deficient isotopes  $^{177-193}$Hg are presented in Fig.~\ref{fig4}. The upper panel of the Fig.~\ref{fig4} shows the comparison of measured and calculated (DD-ME2 oblate, DD-ME2 prolate, DD-PC1 oblate and DD-PC1 prolate) half lives, while the lower panel presents the comparison of measured half-lives with the FRDM \cite{Mol12},  and ground state shapes predicted using the DD-ME2, DD-PC1 and SLy4 interactions for $^{177-193}$Hg. The FRDM calculations were based on the finite-range droplet macroscopic and the folded-Yukawa single-particle microscopic nuclear-structure models. The authors calculated the atomic mass excesses and binding energies, ground-state shell-plus-pairing corrections, ground-state
microscopic corrections, and nuclear ground-state deformations of 9318 nuclei ranging from A = 16 to 339. The standard deviations of the various calculations from the measured data are shown in Table.~\ref{Table 4}. From this table it may be concluded that the half-lives using the deformation values from the FRDM gives the best predictions and is followed by DD-PC1 interaction results. It is further noted that the calculated half-life values do change once the shape of nucleus changes. This may be explained by the fact that our computed GT strength distributions get altered with the deformation values as shown earlier.

\section{Conclusions} \label{sec:conclusions}
In this article, we investigated the effect of deformations on calculated GT strength distributions for neutron deficient  $^{177-193}$Hg isotopes. The nuclear deformation parameters were computed by using the RMF model. The computed deformations were used for GT strength calculations using the deformed pn-QRPA model. Our results  show that the GT strength distributions change with evolving nuclear shapes.  Our findings reflect the sensitivity of computed GT strength distributions and half-lives to nuclear deformations. The summed GT strengths, however, do not react much to the deformation values.

The RMF calculation shows that the nuclei  $^{177-188}$Hg prefer prolate shapes while $^{189-193}$Hg exhibit oblate shapes. On the other hand, the pn-QRPA calculation using SLy4 interaction \cite{Boi15} deduced prolate shapes for $^{177-182}$Hg and oblate shapes for $^{184-196}$Hg. This difference may be traced to the energy profiles within the two models which led to changes in the absolute minima from one deformation to other. (Tuncay can comment further.)  It is to be noted that only even-even Hg isotopes were considered in the SLy4 calculation. Our predicted half-life values for Hg isotopes are in overall better agreement with the measured data as compared to the pn-QRPA calculation of \cite{Boi15} using the SLy4 interaction.

In future measurements of transfer reaction data (single and multinucleon), $B(E2), \rho^2(E0)$ may help theoretical models to better comprehend the physics of shape coexistence.

%

%
%




\begin{table}[ht]
	\scriptsize \caption{Calculated nuclear deformations  for $^{177-193}$Hg. Second, third and fourth columns present the RMF calculation employing the DD-ME2 channel whereas the  fifth, sixth and seventh  columns  present the RMF calculation by DD-PC1 channel. The last column shows the deformation calculated using the FRDM model \cite{Mol12}.}
	\label{Table 1}
	\begin{center}
			\begin{tabular} {c|ccc|ccc|c}
				\hline \\
				Isotopes & G.S. & Prolate & Oblate & G.S.&  Prolate & Oblate &FRDM \\ & DD-ME2 &DD-ME2 & DD-ME2 & DD-PC1&  DD-PC1 & DD-PC1&\\
				\hline
				$^{177}$Hg  & 0.2937 & 0.2937 & -0.3320 & 0.2962 & 0.2962 & -0.3339&-0.115 \\
				$^{179}$Hg  & 0.3078 & 0.3078 & -0.3280 & 0.3154 & 0.3154 & -0.3307&-0.135 \\
				$^{180}$Hg  & 0.3063 & 0.3063 & -0.3259 & 0.3158 & 0.3158 & -0.3290&-0.135 \\
				$^{181}$Hg  & 0.3046 & 0.3046 & -0.3227 & 0.3142 & 0.3142 & -0.3266&-0.146 \\
				$^{182}$Hg  & 0.3045 & 0.3045 & -0.3185 & 0.3125 & 0.3125 & -0.3234&-0.146 \\
				$^{183}$Hg  & 0.3060 & 0.3060 & -0.2098 & 0.3116 & 0.3116 & -0.3198&-0.146 \\
				$^{184}$Hg  & 0.3095 & 0.3095 & -0.2153 & 0.3106 & 0.3106 & -0.2109&-0.146 \\
				$^{185}$Hg  & 0.3003 & 0.3003 & -0.2198 & 0.3037 & 0.3037 & -0.2155&-0.146 \\
				$^{186}$Hg  & 0.2937 & 0.2937 & -0.2230 & 0.2969 & 0.2969 & -0.2179&-0.146\\
				$^{187}$Hg  & 0.2896 & 0.2896 & -0.2198 & 0.2924 & 0.2924 & -0.2140&-0.146 \\
				$^{188}$Hg  & 0.2867 & 0.2867 & -0.1434 & 0.2881 & 0.2881 & -0.1867&-0.146 \\
				$^{189}$Hg  & -0.1424 & 0.2793 & -0.1424 & -0.1436 & 0.2807 & -0.1436&-0.146 \\
				$^{190}$Hg  & -0.1414 & 0.2710 & -0.1414 & -0.1421 & 0.2718 & -0.1421&-0.125 \\
				$^{191}$Hg  & -0.1402 & 0.2613 & -0.1402 & -0.1405 & 0.2605 & -0.1405&-0.125\\
				$^{192}$Hg  & -0.1389 & 0.1066 & -0.1389 & -0.1388 & 0.1103 & -0.1388&-0.125 \\
				$^{193}$Hg  & -0.1373 & 0.1037 & -0.1373 & -0.1368 & 0.1075 & -0.1368&-0.125 \\
				\hline
			\end{tabular}
	\end{center}
\end{table}

\begin{table}[ht]
	\scriptsize \caption{Centroids of calculated GT strength distributions using various interactions for odd-A (upper panel) and even-A (lower panel) Hg isotopes. All values are given in units of $MeV$.}
	\label{Table 2}
	\begin{center}
		\begin{tabular} {ccccccc}
			\hline \\
			Isotopes & DD-ME2&DD-ME2 & DD-ME2 &DD-PC1&DD-PC1 & DD-PC1 \\
			         &G.S.& Oblate & Prolate&G.S. &Oblate & Prolate\\
			\hline
			$^{177}$Hg&1.24& 1.24  & 4.92  &3.94 &3.87  & 4.87 \\
			$^{179}$Hg&1.22& 1.22  & 3.06  &3.02 &1.22  & 3.02 \\
			$^{181}$Hg&2.38& 2.38  & 3.09  &2.86 &2.26  & 3.13 \\
			$^{183}$Hg&1.35& 1.35  & 3.57  &3.47 &0.99  & 3.61 \\
			$^{185}$Hg&1.73& 1.73  & 2.49  &2.52 &1.72  & 2.52 \\
			$^{187}$Hg&1.35& 1.35  & 2.41  & 2.51&1.39  & 2.51 \\
			$^{189}$Hg&1.64& 1.64  & 1.83  & 1.62&1.63  & 1.85 \\
			$^{191}$Hg&0.83& 0.83  & 3.39  &0.83 &0.83  & 3.37 \\
			$^{193}$Hg&2.10& 2.10  & 3.23  & 1.67&2.03  & 3.90 \\
			\hline
			$^{180}$Hg&4.09& 4.09 & 4.57 &4.51&4.02 & 4.51 \\
			$^{182}$Hg&3.11& 3.11 & 3.29 &3.50&3.12 & 3.31 \\
			$^{184}$Hg&4.42& 4.42 & 4.69 & 4.27&4.35 & 4.27\\
			$^{186}$Hg&4.28& 4.28 & 4.47 &4.49 &4.09 & 4.48 \\
			$^{188}$Hg&4.25& 4.25 & 4.33 & 4.32&4.25 & 4.32 \\
			$^{190}$Hg&4.54& 4.54 & 4.79 & 4.41&4.41 & 4.76 \\
			$^{192}$Hg&3.65& 3.65 & 3.99 &3.65 &3.65 & 3.96 \\
			\hline
		\end{tabular}
	\end{center}
\end{table}

\begin{table}[ht]
	\scriptsize \caption{Total strengths of calculated GT strength distributions using various interactions for even-A (upper panel) and odd-A (lower panel) Hg isotopes.}
	\label{Table 3}
	\begin{center}
		\begin{tabular} {ccccccc}
			\hline \\
			Isotopes & DD-ME2&DD-ME2 & DD-ME2 &DD-PC1&DD-PC1 & DD-PC1 \\
			&G.S.& Oblate & Prolate&G.S. &Oblate & Prolate\\
		
		\hline
	$^{180}$Hg & 3.35  & 4.23  & 3.35  & 3.61  & 3.98  & 3.61\\
	$^{182}$Hg & 3.19  & 3.88  & 3.19  & 3.37  & 3.81  & 3.30 \\
    $^{184}$Hg & 3.42  & 3.78  & 3.42  & 3.11  & 3.43  & 3.11 \\
	$^{186}$Hg & 3.91  & 4.12  & 3.91  & 3.78  & 3.78  &3.78 \\
	$^{188}$Hg & 3.29  & 3.80  & 3.29  & 3.45  & 3.85  & 3.45 \\
	$^{190}$Hg& 3.41  & 3.52  & 3.26  & 3.50  & 3.50  & 3.56 \\
	$^{192}$Hg  & 7.86  & 7.86  & 7.86  & 7.85  & 7.85  & 7.71\\
		\hline
	$^{177}$Hg & 1.50  & 0.54  & 1.50  & 1.49  & 28.91 & 1.49\\
	$^{179}$Hg & 0.30  & 0.36  & 0.30  & 0.30  & 0.37  & 0.30\\
	$^{181}$Hg  & 0.03  & 0.10  & 0.03  & 0.04  & 0.17  & 0.04 \\
	$^{183}$Hg  & 0.06  & 0.11  & 0.06  & 0.06  & 0.14  & 0.06 \\
	$^{185}$Hg   & 0.17  & 0.65  & 0.17  & 0.18  & 0.69  & 0.18 \\
	$^{187}$Hg  & 0.18  & 0.15  & 0.18  & 0.17  & 0.13  & 0.17 \\
	$^{189}$Hg  & 0.08  & 0.08  & 0.05  & 0.08  & 0.08  & 0.05 \\
	$^{191}$Hg   & 0.36  & 0.36  & 0.05  & 0.36  & 0.36  & 0.05 \\
	$^{193}$Hg  & 0.03  & 0.03  & 0.03  & 0.03  & 0.03  & 0.02 \\
		\hline
		\end{tabular}
	\end{center}
\end{table}

\begin{table}[ht]
	\scriptsize \caption{Calculated standard deviations from the measured half-lives for the different calculations considered in this work.}
	\label{Table 4}
	\begin{center}
		\begin{tabular} {lc}
			\hline \\
			Calculations & Standard Deviation ($s$) \\
			\hline
			FRDM & 0.60\\
			DD-ME2 Prolate & 4.60\\
			DD-ME2 Oblate & 1.53\\
			DD-ME2 G.S. & 1.45\\
			DD-PC1 Prolate & 10.87\\
			DD-PC1 Oblate & 1.49\\
			DD-PC1 G.S. & 1.40\\
			SLy4 Prolate & 2.33\\
			SLy4 Oblate & 2.61\\	
			SLy4 G.S. & 3.36\\
			\hline
		\end{tabular}
	\end{center}
\end{table}

\clearpage

\end{document}